%% file: BasicProperties-CS.tex
\documentclass[12pt,a4paper]{article}
\usepackage{amsmath}
\usepackage{latexsym}
\makeatletter
\def\rddots{\mathinner{\mkern1mu\raise\p@%
    \vbox{\kern7\p@\hbox{.}}\mkern2mu%
    \raise4\p@\hbox{.}\mkern2mu\raise7\p@\hbox{.}\mkern1mu}}
\makeatother
\makeatletter
   
   \@addtoreset{equation}{section}
\makeatother
\newcommand{\ket}[1]{{\vert{#1}\rangle}}
\newcommand{\bra}[1]{{\langle{#1}\vert}}

\newcommand{\fukuso}{{\mathbf C}}
\newcommand{\real}{{\mathbf R}}

\begin{document}

\title{\sl Basic Properties of Coherent--Squeezed States Revisited}
\author{
  Kazuyuki FUJII
  \thanks{E-mail address : fujii@yokohama-cu.ac.jp }\quad and\ \ 
  Hiroshi Oike
  \thanks{E-mail address : oike@tea.ocn.ne.jp}\\
  ${}^{*}$International College of Arts and Sciences\\
  Yokohama City University\\
  Yokohama, 236--0027\\
  Japan\\
  ${}^{\dagger}$Takado\ 85--5,\ Yamagata, 990--2464\\
  Japan\\
  }
\date{}
\maketitle
\begin{abstract}
  In this paper we treat coherent-squeezed states of 
Fock space once more and study some basic properties 
of them from a geometrical point of view.

  Since the set of coherent-squeezed states $\{\ket{\alpha, \beta}\ 
|\ \alpha, \beta \in \fukuso\}$ makes a real 4-dimensional 
surface in the Fock space ${\cal F}$ (which is of course not flat), 
we can calculate its metric. 

  On the other hand, we know that coherent-squeezed states satisfy 
the minimal uncertainty of Heisenberg under some condition imposed 
on the parameter space $\{\alpha, \beta\}$, so that we can study 
the metric from the view point of uncertainty principle. Then we obtain 
a surprising simple form (at least to us).

  We also make a brief review on Holonomic Quantum Computation 
by use of a simple model based on nonlinear Kerr effect and 
coherent-squeezed operators.
\end{abstract}

\vspace{5mm}\noindent

\vspace{5mm}\noindent

\section{Introduction}
We start with general theory of quantum harmonic oscillator. 
For example see \cite{PD}, \cite{HG} for a standard textbook 
of Quantum Mechanics. 
Let $\{a^{\dagger},a,N\equiv a^{\dagger}a\}$ be a generator 
of the Heisenberg algebra whose relations are given by
\begin{equation}
\label{eq:Heisenberg algebra}
[N,a^{\dagger}]=a^{\dagger},\quad [N,a]=-a,\quad [a,a^{\dagger}]={\bf 1}
\end{equation}
where ${\bf 1}$ is the identity. For the vacuum $\ket{0}$ given by 
\begin{equation}
\label{eq:vacuum}
a\ket{0}=0
\end{equation}
we can define the $n$-th state by
\begin{equation}
\label{eq:n-state}
\ket{n}=\frac{(a^{\dagger})^{n}}{\sqrt{n!}}\ket{0}
\end{equation}
for $n\geq 0$. Then it is easy to see
\begin{equation}
\label{eq:fundamental equations}
a^{\dagger}\ket{n}=\sqrt{n+1}\ket{n+1}, \quad
a\ket{n}=\sqrt{n}\ket{n-1}, \quad
N\ket{n}=n\ket{n}.
\end{equation}
Moreover, we can prove both the orthogonality condition 
and the resolution of unity 
\begin{equation}
\label{eq:resolution of unity}
\langle m|n\rangle=\delta_{mn},\quad
\sum_{n=0}^{\infty}|n\rangle \langle n|={\bf 1}.
\end{equation}
Therefore, we can define the Fock space
\begin{equation}
\label{eq:Fock space}
{\cal F}=\left\{\sum_{n=0}^{\infty}c_{n}\ket{n}\ \in\ \fukuso^{\infty}
\ |\ 
\sum_{n=0}^{\infty}|c_{n}|^{2}<\infty \right\},
\end{equation}
which is a kind of Hilbert space in the mathematical sense. 
Then the inner product of the space is given by
\begin{equation}
\label{eq:Inner product}
\left<\sum_{n=0}^{\infty}c_{n}\ket{n}\ |\sum_{n=0}^{\infty}d_{n}\ket{n}\right>
=
\sum_{n=0}^{\infty}\bar{c}_{n}d_{n}.
\end{equation}

On this space we obtain matrix representations of 
$\{a^{\dagger},a,N\}$ like

\begin{eqnarray}
\label{eq:creation-annihilation}
&&a=
\left(
\begin{array}{ccccc}
0 & 1 &          &          &           \\
  & 0 & \sqrt{2} &          &         \\
  &   & 0        & \sqrt{3} &         \\
  &   &          & 0        & \ddots \\
  &   &          &          & \ddots
\end{array}
\right),\quad
a^{\dagger}=
\left(
\begin{array}{ccccc}
0 &          &          &        &            \\
1 & 0        &          &        &           \\
  & \sqrt{2} & 0        &        &          \\
  &          & \sqrt{3} & 0      &          \\
  &          &          & \ddots & \ddots
\end{array}
\right),
\nonumber \\
&&N=a^{\dagger}a=
\left(
\begin{array}{ccccc}
0 &   &   &   &          \\
  & 1 &   &   &          \\
  &   & 2 &   &          \\
  &   &   & 3 &          \\
  &   &   &   & \ddots
\end{array}
\right)
\end{eqnarray}
by use of (\ref{eq:fundamental equations}).

Now, let us define coherent states and squeezed states which 
play an important role in Quantum Optics in the following.  
For details see for example \cite{WS}, \cite{WM}. 

For $\alpha\in \fukuso$ a coherent state $\ket{\alpha}$ 
is defined by the equation
\begin{equation}
\label{eq:coherent state-1}
a\ket{\alpha}=\alpha\ket{\alpha}.
\end{equation}
The state is rewritten as
\begin{equation}
\label{eq:coherent state-2}
\ket{\alpha}=e^{\alpha a^{\dagger}-\bar{\alpha}a}\ket{0}.
\end{equation}
This expression is in our opinion better than (\ref{eq:coherent state-1}).

Next, if we define operators
\begin{equation}
\label{eq:su(1,1) generator}
K_{+}=\frac{1}{2}(a^{\dagger})^{2},\quad 
K_{-}=\frac{1}{2}a^{2},\quad
K_{3}=\frac{1}{2}\left(N+\frac{1}{2}\right)
\end{equation}
then it is not difficult to see
\begin{equation}
\label{eq:su(1,1) relations}
[K_{3},K_{+}]=K_{+},\quad 
[K_{3},K_{-}]=-K_{-},\quad 
[K_{+},K_{-}]=-2K_{3}.
\end{equation}
This is called the $su(1,1)$ algebra, see for example \cite{KF1}.

For $\beta\in \fukuso$ a squeezed state $\ket{\beta}$ 
is defined by
\begin{equation}
\label{eq:squeezed state}
\ket{\beta}=e^{\beta K_{+}-\bar{\beta}K_{-}}\ket{0}
\end{equation}
which corresponds to (\ref{eq:coherent state-2}) not 
(\ref{eq:coherent state-1}). 

As a result, a coherent-squeezed state is defined as
\begin{equation}
\label{eq:coherent-squeezed state}
\ket{\alpha, \beta}=
e^{\beta K_{+}-\bar{\beta}K_{-}}e^{\alpha a^{\dagger}-\bar{\alpha}a}\ket{0}
\end{equation}
for $\alpha\in \fukuso$ and $\beta\in \fukuso$ (we use the notation 
$\ket{\alpha, \beta}$ in place of $\ket{(\alpha, \beta)}$ for 
simplicity). We set $CS$ a set of all coherent-squeezed states like
\begin{equation}
\label{eq:CS}
CS=\{\ket{\alpha, \beta}\in {\cal F}\ |\ \alpha, \beta \in \fukuso\}.
\end{equation}
This set (manifold) is our target in the paper and we want to study 
some basic properties from a geometrical point of view.

In last in this section, we treat the uncertainty principle by 
Heisenberg.  The position operator $\hat{q}$ and momentum 
operator $\hat{p}$ are expressed as
\begin{equation}
\label{eq:q and p}
\hat{q}=\sqrt{\frac{\hbar}{2\omega}}(a^{\dagger}+a),\quad
\hat{p}=i\sqrt{\frac{\omega\hbar}{2}}(a^{\dagger}-a)
\end{equation}
in terms of $\{a^{\dagger},a\}$. Then the variances of $\hat{q}$ and 
$\hat{p}$ are defined by
\begin{equation}
\label{eq:variances}
(\Delta q)^{2}=\bra{\chi}\hat{q}^{2}\ket{\chi}-\bra{\chi}\hat{q}\ket{\chi}^{2},
\quad
(\Delta p)^{2}=\bra{\chi}\hat{p}^{2}\ket{\chi}-\bra{\chi}\hat{p}\ket{\chi}^{2}
\end{equation}
for any normalized $\ket{\chi} \in {\cal F}$ (we write $\Delta q$ in place of 
$(\Delta q)_{\chi}$ for simplicity) and the uncertainty principle is given by
\begin{equation}
\label{uncertainty principle}
(\Delta q)(\Delta p)\geq \frac{\hbar}{2}.
\end{equation}

We are interested in a state $\ket{\chi}$ giving the minimal value
\begin{equation}
\label{minimal value}
(\Delta q)(\Delta p)=\frac{\hbar}{2}.
\end{equation}
In the next section we set $\ket{\chi}=\ket{\alpha, \beta}$ and 
look for $\{\alpha, \beta\}$ giving the minimal uncertainty 
(\ref{minimal value}).

\section{Coherent-Squeezed States and Uncertainty Principle}
In this section we prove that the coherent-squeezed state 
$\ket{\alpha, \beta}$ (\ref{eq:coherent-squeezed state}) 
gives the minimal uncertainty (\ref{minimal value}) if we choose 
$\alpha\in \fukuso$ and $\beta\in \real$. However, this is 
well-known, see for example \cite{WM}.

From (\ref{eq:variances}) we must calculate the variances 
of $\hat{q}$ and $\hat{p}$ respectively. The result is
\begin{eqnarray}
\label{eq:calculation of variances}
(\Delta q)^{2}
&=&\frac{\hbar}{2\omega}
\left(\cosh|\beta|+\beta\frac{\sinh|\beta|}{|\beta|}\right)
\left(\cosh|\beta|+\bar{\beta}\frac{\sinh|\beta|}{|\beta|}\right), 
\nonumber \\
(\Delta p)^{2}
&=&\frac{\hbar\omega}{2}
\left(\cosh|\beta|-\beta\frac{\sinh|\beta|}{|\beta|}\right)
\left(\cosh|\beta|-\bar{\beta}\frac{\sinh|\beta|}{|\beta|}\right).
\end{eqnarray}
It is very interesting that these have nothing to do with $\alpha$. 
Some calculation gives
\begin{equation}
(\Delta q)^{2}(\Delta p)^{2}=
\left(\frac{\hbar}{2}\right)^{2}
\left\{1-(\beta-\bar{\beta})^{2}\left(\frac{\sinh{2|\beta|}}{2|\beta|}\right)^{2}\right\}.
\end{equation}
From this we have
\begin{equation}
\label{eq:result}
(\Delta q)^{2}(\Delta p)^{2}=\left(\frac{\hbar}{2}\right)^{2}
\ \Longrightarrow\ 
(\Delta q)(\Delta p)=\frac{\hbar}{2}
\end{equation}
if $\beta=\bar{\beta}$ ($\beta$ is real).

As a minimal uncertainty surface we set
\begin{equation}
\label{eq:special CS}
\widetilde{CS}=\{\ket{\alpha, \beta}\in {\cal F}\ 
|\ \alpha\in \fukuso,\ \beta\in \real\}.
\end{equation}

\section{Inner Product of Coherent-Squeezed States}
In this section we give a formula of inner product of 
coherent-squeezed states. Our method is based on 
\cite{EFS}, \cite{FS} or \cite{KF2} (review paper) \footnote{
The method developed in this section is powerful and 
convenient, see \cite{FS-1}, \cite{FS-2} for a further application.}

The formula is well-known for coherent states. For $\alpha$ and 
$\alpha^{\prime}$ it is easy to see
\begin{equation}
\label{eq:inner-product coherent}
\langle{\alpha}\vert{\alpha^{\prime}}\rangle =
e^{\left\{-\frac{1}{2}|\alpha|^{2}-\frac{1}{2}|\alpha^{\prime}|^{2}+
\bar{\alpha}\alpha^{\prime}\right\}}.
\end{equation}

Therefore, our target is to calculate the inner product (\ref{eq:Inner product})
\begin{equation}
\label{eq:inner product}
\langle{\alpha,\beta}\vert{\alpha^{\prime},\beta^{\prime}}\rangle
\end{equation}
for $\alpha, \alpha^{\prime} \in \fukuso$ and 
$\beta, \beta^{\prime} \in \fukuso$. This is not trivial as 
shown in the following.

The inner product (\ref{eq:inner product}) can be written as
\begin{eqnarray*}
\langle{\alpha,\beta}\vert{\alpha^{\prime},\beta^{\prime}}\rangle
&=&
\bra{0}
e^{-(\alpha a^{\dagger}-\bar{\alpha}a)}
e^{-(\beta K_{+}-\bar{\beta}K_{-})}
e^{\beta^{\prime}K_{+}-\overline{\beta^{\prime}}K_{-}}
e^{\alpha^{\prime}a^{\dagger}-\overline{\alpha^{\prime}}a}
\ket{0} \\
&=&
\bra{\alpha}
e^{-(\beta K_{+}-\bar{\beta}K_{-})}
e^{\beta^{\prime}K_{+}-\overline{\beta^{\prime}}K_{-}}
\ket{\alpha^{\prime}}
\end{eqnarray*}
by use of coherent states (\ref{eq:coherent state-2}). 

First of all, we must calculate the (product) operator
\[
e^{-(\beta K_{+}-\bar{\beta}K_{-})}e^{\beta^{\prime}K_{+}-\overline{\beta^{\prime}}K_{-}}
\]
explicitly. For the purpose let us recall the relations (\ref{eq:su(1,1) relations}) 
once more
\[
[K_{3},K_{+}]=K_{+},\quad 
[K_{3},K_{-}]=-K_{-},\quad 
[K_{+},K_{-}]=-2K_{3},\quad
K_{-}=K_{+}^{\dagger}.
\]
If we set $\{k_{+},k_{-},k_{3}\}$ as
\begin{equation}
\label{eq:su(1,1) basic-generators}
 k_{+} = \left(
        \begin{array}{cc}
               0 & 1 \\
               0 & 0 
        \end{array}
       \right),
 \quad 
 k_{-} = \left(
        \begin{array}{cc}
               0 & 0 \\
              -1 & 0 
        \end{array}
       \right),
 \quad 
 k_{3} = \frac12 
       \left(
        \begin{array}{cc}
               1 &  0 \\
               0 & -1 
        \end{array}
       \right)
\end{equation}
then it is very easy to check the relations
\[
[k_{3},k_{+}]=k_{+},\quad 
[k_{3},k_{-}]=-k_{-},\quad 
[k_{+},k_{-}]=-2k_{3},\quad
k_{-}\ne k_{+}^{\dagger}.
\]
That is, $\{k_{+},k_{-},k_{3}\}$ are generators of the Lie algebra 
$su(1,1)$. We show by $SU(1,1)$ the corresponding Lie 
group, which is a typical noncompact group.

Since $SU(1,1)$ is contained in the special linear group $SL(2;\fukuso)$, 
we assume that there exists an infinite dimensional unitary 
representation 
$\rho : SL(2;\fukuso)\ \longrightarrow\ U({\cal F})$ (group homomorphism) 
satisfying
\[
d\rho(k_{+})=K_{+},\quad d\rho(k_{-})=K_{-},\quad d\rho(k_{3})=K_{3}.
\]
Then we have
\begin{eqnarray*}
\label{eq:assumption}
e^{\gamma K_{+}-\bar{\gamma}K_{-}}
&=&
e^{\gamma d\rho(k_{+})-\bar{\gamma}d\rho(k_{-})}  \\
&=&
e^{d\rho(\gamma k_{+}-\bar{\gamma}k_{-})}  \\
&=&
\rho\left(e^{\gamma k_{+}-\bar{\gamma}k_{-}}\right) 
\quad (\Downarrow\ \mbox{by definition}) \nonumber \\
&\equiv& \rho\left(e^{A}\right)
\end{eqnarray*}
where
\[
A=\gamma k_{+}-\bar{\gamma}k_{-}=
\left(
    \begin{array}{cc}
        0 & \gamma         \\
        \bar{\gamma} & 0 
    \end{array}
\right)
\quad
\Longrightarrow
\quad
e^{A}=
\left(
    \begin{array}{cc}
        \cosh|\gamma| & \gamma\frac{\sinh|\gamma|}{|\gamma|}          \\
        \bar{\gamma}\frac{\sinh|\gamma|}{|\gamma|} &  \cosh|\gamma|
    \end{array}
\right).
\]
Therefore, some calculation gives
\begin{eqnarray*}
e^{-(\beta K_{+}-\bar{\beta}K_{-})}e^{\beta^{\prime}K_{+}-\overline{\beta^{\prime}}K_{-}}
&=&
\rho
\left(
\left(
    \begin{array}{cc}
        \cosh|\beta| & -\beta\frac{\sinh|\beta|}{|\beta|}          \\
        -\bar{\beta}\frac{\sinh|\beta|}{|\beta|} & \cosh|\beta|
    \end{array}
\right)
\right)
\rho
\left(
\left(
    \begin{array}{cc}
        \cosh|\beta^{\prime}| & \beta^{\prime}\frac{\sinh|\beta^{\prime}|}{|\beta^{\prime}|}               \\
        \overline{\beta^{\prime}}\frac{\sinh|\beta^{\prime}|}{|\beta^{\prime}|} & \cosh|\beta^{\prime}|
    \end{array}
\right)
\right)  \\
&=&
\rho
\left(
\left(
    \begin{array}{cc}
        \cosh|\beta| & -\beta\frac{\sinh|\beta|}{|\beta|}          \\
        -\bar{\beta}\frac{\sinh|\beta|}{|\beta|} & \cosh|\beta|
    \end{array}
\right)
\left(
    \begin{array}{cc}
        \cosh|\beta^{\prime}| & \beta^{\prime}\frac{\sinh|\beta^{\prime}|}{|\beta^{\prime}|}               \\
        \overline{\beta^{\prime}}\frac{\sinh|\beta^{\prime}|}{|\beta^{\prime}|} & \cosh|\beta^{\prime}|
    \end{array}
\right)
\right)  \\
&\equiv&
\rho
\left(
\left(
  \begin{array}{cc}
  a & b \\
  c & d
  \end{array}
\right)
\right)
\end{eqnarray*}
where
\begin{eqnarray*}
a\equiv a(\beta,\beta^{\prime})
&=&\cosh|\beta|\cosh|\beta^{\prime}|-
       \frac{\beta\overline{\beta^{\prime}}}{|\beta||\beta^{\prime}|}
       \sinh|\beta|\sinh|\beta^{\prime}|, \\
b\equiv b(\beta,\beta^{\prime})
&=&\frac{\beta^{\prime}}{|\beta^{\prime}|}\cosh|\beta|\sinh|\beta^{\prime}|-
        \frac{\beta}{|\beta|}\sinh|\beta|\cosh|\beta^{\prime}|, \\
c\equiv c(\beta,\beta^{\prime})
&=&\frac{\overline{\beta^{\prime}}}{|\beta^{\prime}|}\cosh|\beta|\sinh|\beta^{\prime}|-
        \frac{\bar{\beta}}{|\beta|}\sinh|\beta|\cosh|\beta^{\prime}|=\bar{b}, \\
d\equiv d(\beta,\beta^{\prime})
&=&\cosh|\beta|\cosh|\beta^{\prime}|-
       \frac{\bar{\beta}\beta^{\prime}}{|\beta||\beta^{\prime}|}
       \sinh|\beta|\sinh|\beta^{\prime}|=\bar{a}
\end{eqnarray*}
because $\rho$ is a group homomorphism. For the latter convenience we 
show
\begin{eqnarray*}
\frac{b}{d}
&=&
\frac{-\frac{\beta}{|\beta|}\tanh|\beta|+\frac{\beta^{\prime}}{|\beta^{\prime}|}\tanh|\beta^{\prime}|}
{1-\frac{\bar{\beta}\beta^{\prime}}{|\beta||\beta^{\prime}|}\tanh|\beta|\tanh|\beta^{\prime}|}, \\
\frac{c}{d}
&=&
\frac{-\frac{\bar{\beta}}{|\beta|}\tanh|\beta|+\frac{\overline{\beta^{\prime}}}{|\beta^{\prime}|}\tanh|\beta^{\prime}|}
{1-\frac{\bar{\beta}\beta^{\prime}}{|\beta||\beta^{\prime}|}\tanh|\beta|\tanh|\beta^{\prime}|}=\frac{\bar{b}}{d}.
\end{eqnarray*}

Next, by use of the Gauss decomposition formula (in $SL(2;{\bf C})$)
\[
  \left(
   \begin{array}{cc}
     a & b \\
     c & d \\
   \end{array}
  \right)
=
  \left(
   \begin{array}{cc}
     1 & \frac{b}{d} \\
     0 & 1
   \end{array}
  \right)
  \left(
   \begin{array}{cc}
     \frac{1}{d} & 0 \\
     0 & d
   \end{array}
  \right)
  \left(
   \begin{array}{cc}
     1 & 0            \\
     \frac{c}{d} & 1
   \end{array}
  \right)
\quad (ad-bc=1)
\]
we obtain
\begin{eqnarray*}
\rho
\left(
 \left(
   \begin{array}{cc}
   a & b \\
   c & d
   \end{array}
 \right)
\right)
&=&
\rho
\left(
  \left(
   \begin{array}{cc}
     1 & \frac{b}{d} \\
     0 & 1
   \end{array}
  \right)
  \left(
   \begin{array}{cc}
     \frac{1}{d} & 0 \\
     0 & d
   \end{array}
  \right)
  \left(
   \begin{array}{cc}
     1 & 0            \\
     \frac{c}{d} & 1
   \end{array}
  \right)
\right) \\
&=&
\rho\left(e^{\frac{b}{d}k_{+}}e^{-2\log dk_{3}}e^{-\frac{c}{d}k_{-}}\right) \\
&=&
\rho\left(e^{\frac{b}{d}k_{+}}\right)
\rho\left(e^{-2\log dk_{3}}\right)
\rho\left(e^{-\frac{c}{d}k_{-}}\right) \\
&=&
e^{\frac{b}{d}K_{+}}e^{-2\log dK_{3}}e^{-\frac{c}{d}K_{-}}
\end{eqnarray*}
where we have used again that $\rho$ is a group homomorphism. 
Therefore, our formula is

\vspace{3mm}\noindent
{\bf Formula}
\begin{equation}
e^{-(\beta K_{+}-\bar{\beta}K_{-})}e^{\beta^{\prime}K_{+}-\overline{\beta^{\prime}}K_{-}}=
e^{\frac{b}{d}K_{+}}e^{-2\log dK_{3}}e^{-\frac{c}{d}K_{-}}.
\end{equation}

\vspace{3mm}
However, the proof of this formula is not complete because 
we assumed that $\rho$ is a group homomorphism \footnote{we 
don't know such an explicit construction}.  
We complete the proof as follows (see section 3.2 of \cite{KF2}). 
For $t\geq 0$ we set
\begin{eqnarray*}
F(t)&=&
e^{-t(\beta K_{+}-\bar{\beta}K_{-})}
e^{t(\beta^{\prime}K_{+}-\overline{\beta^{\prime}}K_{-})}, \\
G(t)&=&e^{f(t)K_{+}}e^{g(t)K_{3}}e^{h(t)K_{-}}
\end{eqnarray*}
where
\begin{eqnarray*}
f(t)&=&\frac{b(t)}{d(t)}=
\frac
{-\frac{\beta}{|\beta|}\tanh(t|\beta|)+\frac{\beta^{\prime}}{|\beta^{\prime}|}\tanh(t|\beta^{\prime}|)}
{1-\frac{\bar{\beta}\beta^{\prime}}{|\beta||\beta^{\prime}|}\tanh(t|\beta|)\tanh(t|\beta^{\prime}|)}, \\
g(t)&=&-2\log d(t)=
-2\log\left(\cosh(t|\beta|)\cosh(t|\beta^{\prime}|)-
       \frac{\bar{\beta}\beta^{\prime}}{|\beta||\beta^{\prime}|}
       \sinh(t|\beta|)\sinh(t|\beta^{\prime}|)\right), \\
h(t)&=&\frac{c(t)}{d(t)}=
\frac
{-\frac{\bar{\beta}}{|\beta|}\tanh(t|\beta|)+\frac{\overline{\beta^{\prime}}}{|\beta^{\prime}|}\tanh(t|\beta^{\prime}|)}
{1-\frac{\bar{\beta}\beta^{\prime}}{|\beta||\beta^{\prime}|}\tanh(t|\beta|)\tanh(t|\beta^{\prime}|)}.
\end{eqnarray*}
Then a hard calculation gives
\[
F(0)=G(0)={\bf 1}\quad \mbox{and}\quad F^{\prime}(t)=G^{\prime}(t),
\]
so that we complete the proof required
\[
F(t)=G(t)\ \Longrightarrow\ F(1)=G(1)\ \Longrightarrow\ \mbox{Formula}.
\]

Let us continue our work. From the formula
\begin{eqnarray*}
\langle{\alpha,\beta}\vert{\alpha^{\prime},\beta^{\prime}}\rangle
&=&
\bra{\alpha}
e^{-(\beta K_{+}-\bar{\beta}K_{-})}
e^{\beta^{\prime}K_{+}-\overline{\beta^{\prime}}K_{-}}
\ket{\alpha^{\prime}} \\
&=&
\bra{\alpha}
e^{\frac{b}{d}K_{+}}e^{-2\log dK_{3}}e^{-\frac{c}{d}K_{-}}
\ket{\alpha^{\prime}} \\
&=&
\bra{\alpha}e^{-2\log dK_{3}}\ket{\alpha^{\prime}}
e^{\frac{b}{d}\frac{\bar{\alpha}^{2}}{2}-
\frac{\bar{b}}{d}\frac{{\alpha^{\prime}}^{2}}{2}} \\
&=&
\bra{\alpha}e^{-\log d N}\ket{\alpha^{\prime}}
e^{\frac{b}{d}\frac{\bar{\alpha}^{2}}{2}-
\frac{\bar{b}}{d}\frac{{\alpha^{\prime}}^{2}}{2}-\frac{1}{2}\log d}
\end{eqnarray*}
where we have used equations
\[
K_{-}\ket{\alpha^{\prime}}=\frac{1}{2}a^{2}\ket{\alpha^{\prime}}=
\frac{{\alpha^{\prime}}^{2}}{2}\ket{\alpha^{\prime}},\ \ 
\bra{\alpha}K_{+}=\frac{\bar{\alpha}^{2}}{2}\bra{\alpha},\ \ 
K_{3}=\frac{1}{2}(N+\frac{1}{2}).
\]
Therefore, we have only to calculate the term 
$\bra{\alpha}e^{-\log d N}\ket{\alpha^{\prime}}$. For the purpose 
we set
\[
f(t)=\bra{\alpha}e^{-t\log d N}\ket{\alpha^{\prime}}
\]
for $t\geq 0$. It is $f(0)=\langle{\alpha}|{\alpha^{\prime}}\rangle$ and
\begin{eqnarray*}
f^{\prime}(t)
&=&-\log d\ \bra{\alpha}e^{-t\log d N}N\ket{\alpha^{\prime}} \\
&=&-\log d\ 
\bra{\alpha}e^{-t\log d N}a^{\dagger}\ket{\alpha^{\prime}}\alpha^{\prime} \\
&=&-\log d\ 
\bra{\alpha}e^{-t\log d N}a^{\dagger}e^{t\log d N}e^{-t\log d N}
\ket{\alpha^{\prime}}\alpha^{\prime} \\
&=&-\log d\ 
\bra{\alpha}e^{-t\log d}a^{\dagger}e^{-t\log d N}\ket{\alpha^{\prime}}\alpha^{\prime} \\
&=&-\log d\ e^{-t\log d}\bar{\alpha}\alpha^{\prime}\bra{\alpha}e^{-t\log d N}\ket{\alpha^{\prime}} \\
&=&-\log d\ e^{-t\log d}\bar{\alpha}\alpha^{\prime}f(t).
\end{eqnarray*}
That is, we have 
\[
f(0)=\langle{\alpha}|{\alpha^{\prime}}\rangle\quad \mbox{and}\quad
f^{\prime}(t)=-\log d\ e^{-t\log d}\bar{\alpha}\alpha^{\prime}f(t).
\]
The solution is given by
\[
f(t)=\langle{\alpha}|{\alpha^{\prime}}\rangle
e^{\bar{\alpha}\alpha^{\prime}\left(e^{-t\log d}-1\right)}
\]
and we have
\[
f(1)
=\langle{\alpha}|{\alpha^{\prime}}\rangle
e^{\bar{\alpha}\alpha^{\prime}\left(e^{-\log d}-1\right)}
=\langle{\alpha}|{\alpha^{\prime}}\rangle
e^{\bar{\alpha}\alpha^{\prime}\left(\frac{1}{d}-1\right)}.
\]

As a result we obtain
\begin{eqnarray*}
\langle{\alpha,\beta}\vert{\alpha^{\prime},\beta^{\prime}}\rangle
&=&
\langle{\alpha}|{\alpha^{\prime}}\rangle
e^{
\left(\frac{1}{d}-1\right)\bar{\alpha}\alpha^{\prime}+
\frac{b}{d}\frac{\bar{\alpha}^{2}}{2}-
\frac{\bar{b}}{d}\frac{{\alpha^{\prime}}^{2}}{2}-\frac{1}{2}\log d
}\\
&=&
\langle{\alpha}|{\alpha^{\prime}}\rangle\ \frac{1}{\sqrt{d}}
e^{
\frac{b}{d}\frac{\bar{\alpha}^{2}}{2}-
\frac{\bar{b}}{d}\frac{{\alpha^{\prime}}^{2}}{2}+
\left(\frac{1}{d}-1\right)\bar{\alpha}\alpha^{\prime}
}\\
&=&
\langle{\alpha}|{\alpha^{\prime}}\rangle
\langle{\beta}|{\beta^{\prime}}\rangle
e^{
\frac{b}{d}\frac{\bar{\alpha}^{2}}{2}-
\frac{\bar{b}}{d}\frac{{\alpha^{\prime}}^{2}}{2}+
\left(\frac{1}{d}-1\right)\bar{\alpha}\alpha^{\prime}
}
\end{eqnarray*}
because
\begin{equation}
\label{eq:inner-product squeezed}
\langle{\beta}|{\beta^{\prime}}\rangle
=\frac{1}{\sqrt{d}}
=\frac{1}{\sqrt{\cosh|\beta|\cosh|\beta^{\prime}|-
                    \frac{\bar{\beta}\beta^{\prime}}{|\beta||\beta^{\prime}|}
                    \sinh|\beta|\sinh|\beta^{\prime}|}}.
\end{equation}

\vspace{3mm}\noindent
{Note}\ \ With respect to $d$ it may be better to rewrite as
\[
d=\cosh|\beta|\cosh|\beta^{\prime}|
\left\{1-\frac{\bar{\beta}\beta^{\prime}}{|\beta||\beta^{\prime}|}
\tanh|\beta|\tanh|\beta^{\prime}|\right\}.
\]

\vspace{3mm}
Let us summarize the result as

\noindent
{\bf Theorem I}
\begin{equation}
\label{eq:main result}
\langle{\alpha,\beta}\vert{\alpha^{\prime},\beta^{\prime}}\rangle
=
\langle{\alpha}|{\alpha^{\prime}}\rangle
\langle{\beta}|{\beta^{\prime}}\rangle
\exp\left\{
\frac{b}{d}\frac{\bar{\alpha}^{2}}{2}-
\frac{\bar{b}}{d}\frac{{\alpha^{\prime}}^{2}}{2}+
\left(\frac{1}{d}-1\right)\bar{\alpha}\alpha^{\prime}
\right\}.
\end{equation}

This is an interesting formula and has not been given 
as far as we know.

\section{Metric of Coherent-Squeezed States}
In this section we determine a metric induced from 
the inner product of the Fock space ${\cal F}$  
for the surface $CS$ in (\ref{eq:CS}) and $\widetilde{CS}$ in 
(\ref{eq:special CS}). 
Let us cite some necessary notations from \cite{PV}. 

We set $\varphi$ a set of parameters 
$(\varphi_{1},\varphi_{2},\cdots, \varphi_{k})$ and 
$\ket{\varphi}$ a normalized vector in ${\cal F}$ in (\ref{eq:Fock space}). 
Assuming that the parameter space is flat, we set a difference vector 
as
\[
\delta\ket{\varphi}=\ket{\varphi+d\varphi}-\ket{\varphi}
\quad
(d\varphi=(d\varphi_{1}, d\varphi_{2}, \cdots, d\varphi_{k})).
\]
Then the (induced) metric $d l^{2}$ is given by
\begin{equation}
\label{eq:metric}
d l^{2}=
\langle{\delta\ket{\varphi}}|{\delta\ket{\varphi}}\rangle-
|\langle{\ket{\varphi}}|{\delta\ket{\varphi}}\rangle|^{2},
\end{equation}
where $\langle{\cdot}|{\cdot}\rangle$ is the inner product in 
(\ref{eq:inner product}). See (2.13) in \cite{PV} for details.

Some algebra gives
\begin{equation}
\label{eq:metric 2}
d l^{2}
=1-|\langle{\ket{\varphi}} | {\ket{\varphi+d\varphi}}\rangle|^{2}
=1-|\langle{\varphi}|{\varphi+d\varphi}\rangle|^{2},
\end{equation}
so we have only to calculate the term 
$\langle{\varphi}|{\varphi+d\varphi}\rangle\equiv
\langle{\varphi}||{\varphi+d\varphi}\rangle\ (\mbox{bra-ket})$.

Let us list our results (we omit the calculation because it is 
very complicated).

\subsection{Coherent States}
This case is very easy. From (\ref{eq:inner-product coherent}) 
the result is simply
\begin{equation}
\label{eq:metric of CS}
dl^{2}
=1-|\langle{\alpha}|{\alpha+d\alpha}\rangle|^{2}
=d\alpha d\bar{\alpha}=(d\alpha_{1})^{2}+(d\alpha_{2})^{2}
\end{equation}
if we write $d\alpha=d\alpha_{1}+i d\alpha_{2}\ (d\bar{\alpha}=
d\alpha_{1}-i d\alpha_{2})$. See \S{2} in \cite{PV}.

\subsection{Squeezed States}
This case is not easy. First, we introduce the notation
\[
K=\frac{\cosh(|\beta|)\sinh(|\beta|)}{|\beta|}=\frac{\sinh(2|\beta|)}{2|\beta|}
\]
for later convenience. We must calculate
\[
dl^{2}=1-|\langle{\beta}|{\beta+d\beta}\rangle|^{2}
\]
and from (\ref{eq:inner-product squeezed}) the result is
\begin{eqnarray}
\label{eq:metric of SS}
dl^{2}
&=&
\frac{1}{8}\frac{\bar{\beta}^{2}}{|\beta|^{2}}(1-K^{2})(d\beta)^{2}+
\frac{2}{8}(1+K^{2}){d\beta}{d\bar{\beta}}+
\frac{1}{8}\frac{{\beta}^{2}}{|\beta|^{2}}(1-K^{2})(d\bar{\beta})^{2} 
\nonumber \\
&=&
\frac{1}{8}
\left\{2(1+K^{2})+\left(\frac{{\beta}^{2}}{|\beta|^{2}}+
\frac{\bar{\beta}^{2}}{|\beta|^{2}}\right)(1-K^{2})\right\}(d\beta_{1})^{2}- \nonumber \\
&{}&
\frac{2i}{8}\left(\frac{{\beta}^{2}}{|\beta|^{2}}-\frac{\bar{\beta}^{2}}{|\beta|^{2}}\right)
(1-K^{2})d\beta_{1}d\beta_{2}+ \nonumber \\
&{}&
\frac{1}{8}
\left\{2(1+K^{2})-\left(\frac{{\beta}^{2}}{|\beta|^{2}}+
\frac{\bar{\beta}^{2}}{|\beta|^{2}}\right)(1-K^{2})\right\}(d\beta_{2})^{2}
\end{eqnarray}
if we write $d\beta=d\beta_{1}+i d\beta_{2}\ (d\bar{\beta}=d\beta_{1}-i d\beta_{2})$. 
The determinant of the metric is 
\[
|g|=\frac{1}{4}K^{2}\ (>0).
\]

If $\beta$ is real ($\beta=\bar{\beta}\ \Rightarrow\ \beta=\beta_{1}$) 
then we have
\begin{equation}
\label{eq:metric of SS special}
dl^{2}=\frac{1}{2}(d\beta_{1})^{2}.
\end{equation}
That is, the metric is extremely simple.

\subsection{Coherent-Squeezed States}
This case is very and very hard \footnote{We used MATHEMATICA 
in order to calculate the metric. The calculation by ``hand" may be 
almost impossible.}. 
First, we introduce the notations
\[
K_{1}=\frac{\sinh(|\beta|)}{|\beta|},\quad
K_{2}=\frac{\sinh(2|\beta|)}{2|\beta|},\quad
K_{4}=\frac{\sinh(4|\beta|)}{4|\beta|}
\]
and
\begin{eqnarray*}
X&=&\beta-3\bar{\beta}-2\beta\cosh(2|\beta|)+2(\beta+2\bar{\beta})K_{2}-(\beta+\bar{\beta})K_{4}, \\
Y&=&-\beta+\bar{\beta}-2\bar{\beta}K_{2}+(\beta+\bar{\beta})K_{4}, \\
Z&=&\beta+\bar{\beta}+2\beta\cosh(2|\beta|)-2\beta K_{2}-(\beta+\bar{\beta})K_{4}
\end{eqnarray*}
and
\begin{eqnarray*}
F&=&-1+4|\beta|^{2}+2\cosh(2|\beta|)-\cosh(4|\beta|), \\
G&=&\ \ 1+4|\beta|^{2}-2\cosh(2|\beta|)+\cosh(4|\beta|).
\end{eqnarray*}
Let us note down (very) important equations
\begin{equation}
\label{eq:important equations}
X+2Y+Z=0\quad \mbox{and}\quad F+G=8|\beta|^{2}.
\end{equation}

We must calculate
\[
dl^{2}=1-|\langle{\alpha,\beta}\vert{\alpha+d\alpha,\beta+d\beta}\rangle|^{2}
\]
and from (\ref{eq:main result}) the result is
\begin{eqnarray}
\label{eq:metric of CS-S}
dl^{2}
&=&d\alpha d\bar{\alpha}+ \nonumber \\
&{}&\frac{1}{8}\left\{\frac{\bar{\beta}^{2}}{|\beta|^{2}}(1-K_{2}^{2})+
\frac{\overline{Z}\alpha^{2}+F|\alpha|^{2}+X\bar{\alpha}^{2}}{\beta^{2}}\right\}(d\beta)^{2}+ \nonumber \\
&{}&\frac{2}{8}\left\{(1+K_{2}^{2})+
\frac{\overline{Y}\alpha^{2}+G|\alpha|^{2}+Y\bar{\alpha}^{2}}{|\beta|^{2}}\right\}d\beta d\bar{\beta}+ \nonumber \\
&{}&\frac{1}{8}\left\{\frac{\beta^{2}}{|\beta|^{2}}(1-K_{2}^{2})+
\frac{\overline{X}\alpha^{2}+F|\alpha|^{2}+Z\bar{\alpha}^{2}}{\bar{\beta}^{2}}\right\}(d\bar{\beta})^{2}+ \nonumber \\
&{}&
\frac{1}{2}\left\{\frac{\alpha{\bar{\beta}}^{2}}{|\beta|^{2}}(1-K_{2})-
\bar{\alpha}\bar{\beta}K_{1}^{2}\right\}d\alpha d\beta + \nonumber \\
&{}&
\frac{1}{2}\left\{\alpha(1+K_{2})+\bar{\alpha}\beta K_{1}^{2}\right\}d\alpha d\bar{\beta} +  \nonumber \\
&{}&
\frac{1}{2}\left\{\bar{\alpha}(1+K_{2})+\alpha\bar{\beta}K_{1}^{2}\right\}d\bar{\alpha}d\beta + \nonumber \\
&{}&
\frac{1}{2}\left\{\frac{\bar{\alpha}{\beta}^{2}}{|\beta|^{2}}(1-K_{2})-
\alpha\beta K_{1}^{2}\right\}d\bar{\alpha}d\bar{\beta}.
\end{eqnarray}

Let us rewrite this into a real matrix form. By setting 
$d\alpha=d\alpha_{1}+i d\alpha_{2}\ (d\bar{\alpha}=d\alpha_{1}-i d\alpha_{2})$ and 
$d\beta=d\beta_{1}+i d\beta_{2}\ (d\bar{\beta}=d\beta_{1}-i d\beta_{2})$ 
we have

\noindent
{\bf Theorem II}
\begin{equation}
\label{eq:metric of CS-S real form}
dl^{2}=
(d\alpha_{1}, d\alpha_{2}, d\beta_{1}, d\beta_{2})
\left(
    \begin{array}{cccc}
      1       & 0      & g_{13} & g_{14}  \\
      0       & 1      & g_{23} & g_{24}  \\
      g_{13} & g_{23} & g_{33} & g_{34}  \\
      g_{14} & g_{24} & g_{34} & g_{44}
    \end{array}
\right)
\left(
    \begin{array}{c}
     d\alpha_{1} \\
     d\alpha_{2} \\
     d\beta_{1}  \\
     d\beta_{2}
    \end{array}
\right)
\end{equation}
where

\begin{eqnarray}
\label{eq:metric of CS surface}
g_{33}
&=&\frac{1}{8}
\left\{
2(1+K_{2}^{2})+\frac{\beta^{2}+\bar{\beta}^{2}}{|\beta|^{2}}(1-K_{2}^{2})+
2\frac{\overline{Y}\alpha^{2}+G|\alpha|^{2}+Y\bar{\alpha}^{2}}{|\beta|^{2}}
\right. \nonumber \\
&{}&
\left.\quad\ 
+\frac{\overline{Z}\alpha^{2}+F|\alpha|^{2}+X\bar{\alpha}^{2}}{\beta^{2}}+
\frac{\overline{X}\alpha^{2}+F|\alpha|^{2}+Z\bar{\alpha}^{2}}{\bar{\beta}^{2}}
\right\}, \nonumber \\
g_{44}
&=&\frac{1}{8}
\left\{
2(1+K_{2}^{2})-\frac{\beta^{2}+\bar{\beta}^{2}}{|\beta|^{2}}(1-K_{2}^{2})+
2\frac{\overline{Y}\alpha^{2}+G|\alpha|^{2}+Y\bar{\alpha}^{2}}{|\beta|^{2}}
\right. \nonumber \\
&{}&
\left.\quad\ 
-\frac{\overline{Z}\alpha^{2}+F|\alpha|^{2}+X\bar{\alpha}^{2}}{\beta^{2}}-
\frac{\overline{X}\alpha^{2}+F|\alpha|^{2}+Z\bar{\alpha}^{2}}{\bar{\beta}^{2}}
\right\}, \nonumber \\
g_{34}&=&g_{43}=\frac{i}{8}
\left\{
-\frac{\beta^{2}-\bar{\beta}^{2}}{|\beta|^{2}}(1-K_{2}^{2})+
\frac{\overline{Z}\alpha^{2}+F|\alpha|^{2}+X\bar{\alpha}^{2}}{\beta^{2}}-
\frac{\overline{X}\alpha^{2}+F|\alpha|^{2}+Z\bar{\alpha}^{2}}{\bar{\beta}^{2}}
\right\}, \nonumber \\
g_{13}&=&g_{31}=\frac{1}{4}
\left\{
\frac{\alpha{\bar{\beta}}^{2}+\bar{\alpha}\beta^{2}}{|\beta|^{2}}(1-K_{2})+
(\alpha+\bar{\alpha})(1+K_{2})
-(\alpha-\bar{\alpha})(\beta-\bar{\beta})K_{1}^{2}
\right\}, \nonumber \\
g_{24}&=&g_{42}=\frac{1}{4}
\left\{
-\frac{\alpha{\bar{\beta}}^{2}+\bar{\alpha}\beta^{2}}{|\beta|^{2}}(1-K_{2})+
(\alpha+\bar{\alpha})(1+K_{2})
+(\alpha+\bar{\alpha})(\beta+\bar{\beta})K_{1}^{2}
\right\}, \nonumber \\
g_{14}&=&g_{41}=\frac{i}{4}
\left\{
\frac{\alpha{\bar{\beta}}^{2}-\bar{\alpha}\beta^{2}}{|\beta|^{2}}(1-K_{2})
-(\alpha-\bar{\alpha})(1+K_{2})
+(\alpha-\bar{\alpha})(\beta+\bar{\beta})K_{1}^{2}
\right\}, \nonumber \\
g_{23}&=&g_{32}=\frac{i}{4}
\left\{
\frac{\alpha{\bar{\beta}}^{2}-\bar{\alpha}\beta^{2}}{|\beta|^{2}}(1-K_{2})
+(\alpha-\bar{\alpha})(1+K_{2})
+(\alpha+\bar{\alpha})(\beta-\bar{\beta})K_{1}^{2}
\right\}.
\end{eqnarray}

\vspace{3mm}
The determinant of the metric is
\[
|g|
\equiv 
\left|
  \begin{array}{cc}
     E     & A  \\
     A^{t} & B
  \end{array}
\right|
=
|B-A^{t}A|
=
\left|
  \begin{array}{cc}
     g_{33}-({g_{13}}^{2}+{g_{23}}^{2}) & g_{34}-(g_{13}g_{14}+g_{23}g_{24})  \\
     g_{34}-(g_{13}g_{14}+g_{23}g_{24}) & g_{44}-({g_{14}}^{2}+{g_{24}}^{2}) 
  \end{array}
\right|.
\]
In spite of our very effort we could not obtain a compact form 
of $|g|$,  so we present

\vspace{3mm}\noindent
{\bf Problem}\ \ Give a compact form to $|g|$.

\vspace{3mm}
We will report some mathematical experiments by use of 
MATHEMATICA in another paper, \cite{FO}.

\section{Special Case : $\alpha \in \fukuso$ and $\beta \in \real$}
In this section, let us study the metric by restricting to the case 
$\alpha \in \fukuso$ and $\beta \in \real$ (see Section II).

For simplicity we set $\beta=\beta_{1}$. Then, 
the metric (\ref{eq:metric of CS-S real form}) becomes a dramatic  
simple form

\noindent
{\bf Theorem III}
\begin{equation}
\label{eq:metric of CS-S real form (special)}
dl^{2}=
(d\alpha_{1}, d\alpha_{2}, d\beta)
\left(
    \begin{array}{ccc}
      1       & 0      & \frac{\alpha+\bar{\alpha}}{2}                                                         \\
      0       & 1      &  i\frac{\alpha-\bar{\alpha}}{2}                                                        \\
      \frac{\alpha+\bar{\alpha}}{2} & i\frac{\alpha-\bar{\alpha}}{2} & \frac{1+4|\alpha|^{2}}{2}
    \end{array}
\right)
\left(
    \begin{array}{c}
     d\alpha_{1} \\
     d\alpha_{2} \\
     d\beta
    \end{array}
\right)
\end{equation}
by use of the equations (\ref{eq:important equations}). 

It is very interesting (mysterious) that there is no $\beta$-dependence 
on $g=(g_{ij})$. Therefore, the determinant of the metric is simply
\[
|g|=\frac{1+2|\alpha|^{2}}{2}=\frac{1}{2}+|\alpha|^{2}.
\]

\section{Holonomic Quantum Computation}
In this section let us make a brief review of Holonomic Quantum 
Computation for readers by making use of a simple model based 
on coherent-squeezed operators. See for example \cite{ZR}, 
\cite{KF} and \cite{PZR}.

From (\ref{eq:coherent-squeezed state}) we set
\begin{equation}
\label{eq:coherent-squeezed operator}
{\cal O}(\alpha,\beta)=V(\beta)U(\alpha)\equiv 
e^{\beta K_{+}-\bar{\beta}K_{-}}e^{\alpha a^{\dagger}-\bar{\alpha}a}
\end{equation}
for $\alpha,\beta \in \fukuso$. Of course, this operator is unitary 
and called a coherent-squeezed operator in the following.

Next, we define an effective Hamiltonian representing a nonlinear 
Kerr effect
\begin{equation}
\label{eq:nonlinear kerr effect}
H_{0}=\hbar\omega (a^{\dagger})^{2}a^{2}=\hbar\omega N(N-1).
\end{equation}
If we set
\[
{\cal C}=\mbox{Vect}_{\fukuso}\{\ket{0},\ket{1}\}
\]
then we have
\[
H_{0}{\cal C}=O.
\]
The space of eigenstates corresponding to $0$ is two dimensional, 
which will become a qubit space. 
In order to construct a geometric method of Quantum Computation 
we utilize this (degenerate) space.

By use of the operator (\ref{eq:coherent-squeezed operator}) 
we can define a family of Hamiltonians by
\begin{equation}
\label{eq:nonlinear kerr effect rotation}
H_{(\alpha,\beta)}={\cal O}(\alpha,\beta)H_{0}{\cal O}(\alpha,\beta)^{\dagger}.
\end{equation}
Then ${\cal O}(\alpha,\beta){\cal C}$ becomes
\[
H_{(\alpha,\beta)}{\cal O}(\alpha,\beta){\cal C}={\cal O}(\alpha,\beta)H_{0}{\cal C}=O,
\]
so that we have a family of two dimensional vector spaces in 
${\cal F}$ parametrized by $(\alpha,\beta)$. 
This is just a vector bundle over $\fukuso \times \fukuso$ \footnote{
To be precise, it is a little ambiguous, see \cite{KF} for more details.}

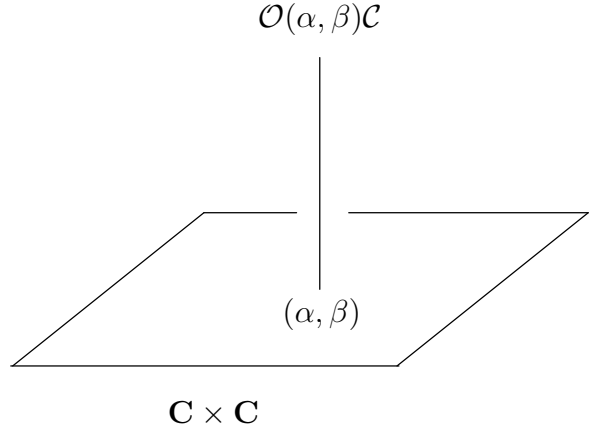
\vspace{3mm}
\begin{center}
\input{vector-bundle.tex}
\end{center}

\vspace{5mm}
From this, a canonical connection form ${\cal A}$ is given by
\begin{equation}
\label{eq:connection}
{\cal A}
={\cal O}(\alpha,\beta)^{\dagger}d{\cal O}(\alpha,\beta)
\equiv {\cal O}^{\dagger}d{\cal O}
\quad (\Longrightarrow {\cal A}^{\dagger}=-{\cal A})
\end{equation}
where
\[
d=d\alpha\frac{\partial}{\partial \alpha}+d\beta\frac{\partial}{\partial \beta}+
d\bar{\alpha}\frac{\partial}{\partial \bar{\alpha}}+
d\bar{\beta}\frac{\partial}{\partial \bar{\beta}},
\]
and therefore the curvature form ${\cal F}$ becomes
\begin{equation}
\label{eq:curvature}
{\cal F}
=d{\cal A}+{\cal A}\wedge{\cal A}
=d{\cal O}^{\dagger}\wedge d{\cal O}+
{\cal O}^{\dagger}d{\cal O}\wedge {\cal O}^{\dagger}d{\cal O}.
\end{equation}

Next, let us define the holonomy operator. Let $\gamma$ be 
a loop in the space $\fukuso\times \fukuso$ starting from 
$(0,0)$ like
\[
\gamma : [0,1]\longrightarrow \fukuso\times\fukuso\ 
\mbox{(differentiable)}, 
\quad \gamma(0)=\gamma(1)=(0,0).
\]
For a loop $\gamma$ the holonomy (operator) $\Gamma$ is 
defined by the path--ordered exponential integral (along $\gamma$) 
like
\begin{equation}
\label{eq:holonomy}
\Gamma(\gamma)={\cal P}\exp\left(\int_{\gamma}{\cal A}\right)\ \in\ U(2).
\end{equation}
See the following figure.
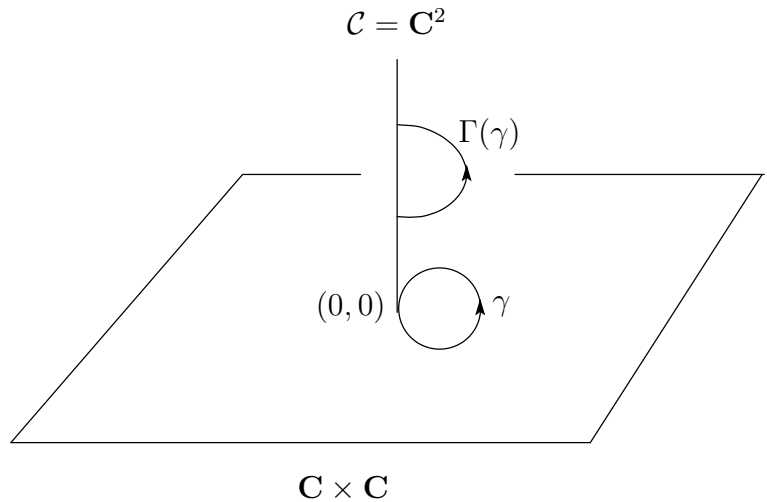
\vspace{3mm}
\begin{center}
\input{holonomy.tex}
\end{center}
\vspace{5mm}

Now we are in a position to calculate these quantities. However, 
we don't repeat them in the paper. 
Read \cite{ZR}, \cite{KF} and \cite{PZR} in detail.

\section{Concluding Remarks}
In the paper we revisited coherent-squeezed states and 
calculated the inner product and gave the metric 
induced from the Fock space (a kind of Hilbert space). 

Its metric form is very complicated but becomes extremely 
simple by considering the special case satisfying the minimal 
uncertainty of Heisenberg. 

We also reviewed a non-abelian holonomic Quantum 
Computation by taking a simple model performed by 
coherent-squeezed operators.

We would like to emphasize that coherent-squeezed states or 
coherent-squeezed operators play a central role in 
Quantum Optics or Mathematical Physics, so that we 
expect young researchers to generalize our results widely.

\vspace{10mm}\noindent 
{\bf Acknowledgments}\\
We would like to thank Ryu Sasaki and Tatsuo Suzuki 
for useful suggestions and comments.

%%%%%%%%%%%%%
%References%
%%%%%%%%%%%%%

\end{document}

%% file: vector-bundle.tex
%WinTpicVersion3.08
\unitlength 0.1in
\begin{picture}( 29.9000, 20.6000)( 20.0000,-23.6500)
% LINE 2 0 3 0
% 2 3000 1400 2010 2200
% 
\special{pn 8}%
\special{pa 3000 1400}%
\special{pa 2010 2200}%
\special{fp}%
% LINE 2 0 3 0
% 2 4990 1400 4000 2200
% 
\special{pn 8}%
\special{pa 4990 1400}%
\special{pa 4000 2200}%
\special{fp}%
% LINE 2 0 3 0
% 2 2000 2200 4010 2200
% 
\special{pn 8}%
\special{pa 2000 2200}%
\special{pa 4010 2200}%
\special{fp}%
% LINE 2 0 3 0
% 2 3600 590 3600 1800
% 
\special{pn 8}%
\special{pa 3600 590}%
\special{pa 3600 1800}%
\special{fp}%
% LINE 2 0 3 0
% 2 3000 1400 3480 1400
% 
\special{pn 8}%
\special{pa 3000 1400}%
\special{pa 3480 1400}%
\special{fp}%
% STR 2 0 3 0
% 3 3050 2350 3050 2450 5 0
% $\fukuso\times \fukuso$
\put(30.5000,-24.5000){\makebox(0,0){$\fukuso\times \fukuso$}}%
% STR 2 0 3 0
% 3 3610 1830 3610 1930 5 0
% $(\alpha,\beta)$
\put(36.1000,-19.3000){\makebox(0,0){$(\alpha,\beta)$}}%
% STR 2 0 3 0
% 3 3600 290 3600 390 5 0
% ${\cal O}(\alpha,\beta){\cal C}$
\put(36.0000,-3.9000){\makebox(0,0){${\cal O}(\alpha,\beta){\cal C}$}}%
% LINE 2 0 3 0
% 2 3750 1400 4990 1400
% 
\special{pn 8}%
\special{pa 3750 1400}%
\special{pa 4990 1400}%
\special{fp}%
\end{picture}%

%% file: holonomy.tex
%WinTpicVersion3.08
\unitlength 0.1in
\begin{picture}( 39.0000, 24.4000)( 16.0000,-27.5500)
% LINE 2 0 3 0
% 2 1600 2600 4600 2600
% 
\special{pn 8}%
\special{pa 1600 2600}%
\special{pa 4600 2600}%
\special{fp}%
% LINE 2 0 3 0
% 2 4210 1200 5500 1200
% 
\special{pn 8}%
\special{pa 4210 1200}%
\special{pa 5500 1200}%
\special{fp}%
% STR 2 0 3 0
% 3 3320 2740 3320 2840 5 0
% $\fukuso\times \fukuso$
\put(33.2000,-28.4000){\makebox(0,0){$\fukuso\times \fukuso$}}%
% STR 2 0 3 0
% 3 3360 1790 3360 1890 5 0
% $(0,0)$
\put(33.6000,-18.9000){\makebox(0,0){$(0,0)$}}%
% STR 2 0 3 0
% 3 3600 300 3600 400 5 0
% ${\cal C}=\fukuso^{2}$
\put(36.0000,-4.0000){\makebox(0,0){${\cal C}=\fukuso^{2}$}}%
% LINE 2 0 3 0
% 2 2800 1200 1600 2600
% 
\special{pn 8}%
\special{pa 2800 1200}%
\special{pa 1600 2600}%
\special{fp}%
% LINE 2 0 3 0
% 2 2800 1200 3410 1200
% 
\special{pn 8}%
\special{pa 2800 1200}%
\special{pa 3410 1200}%
\special{fp}%
% SPLINE 2 0 3 0
% 9 3600 1420 3720 1420 3900 1330 3960 1210 3930 1100 3830 1000 3710 950 3600 940 3600 940
% 
\special{pn 8}%
\special{pa 3600 1420}%
\special{pa 3632 1422}%
\special{pa 3664 1424}%
\special{pa 3696 1422}%
\special{pa 3728 1420}%
\special{pa 3760 1412}%
\special{pa 3790 1402}%
\special{pa 3820 1390}%
\special{pa 3848 1374}%
\special{pa 3876 1354}%
\special{pa 3900 1332}%
\special{pa 3922 1306}%
\special{pa 3940 1278}%
\special{pa 3952 1248}%
\special{pa 3960 1218}%
\special{pa 3960 1186}%
\special{pa 3954 1154}%
\special{pa 3942 1124}%
\special{pa 3928 1094}%
\special{pa 3908 1068}%
\special{pa 3886 1044}%
\special{pa 3862 1024}%
\special{pa 3836 1004}%
\special{pa 3808 988}%
\special{pa 3780 974}%
\special{pa 3750 962}%
\special{pa 3720 952}%
\special{pa 3688 946}%
\special{pa 3656 944}%
\special{pa 3624 942}%
\special{pa 3600 940}%
\special{sp}%
% LINE 2 0 3 0
% 2 5490 1200 4600 2600
% 
\special{pn 8}%
\special{pa 5490 1200}%
\special{pa 4600 2600}%
\special{fp}%
% CIRCLE 2 0 3 0
% 4 3820 1900 4000 2010 4030 1900 4030 1900
% 
\special{pn 8}%
\special{ar 3820 1900 212 212  0.0000000 6.2831853}%
% VECTOR 2 0 3 0
% 2 4030 1890 4030 1860
% 
\special{pn 8}%
\special{pa 4030 1890}%
\special{pa 4030 1860}%
\special{fp}%
\special{sh 1}%
\special{pa 4030 1860}%
\special{pa 4010 1928}%
\special{pa 4030 1914}%
\special{pa 4050 1928}%
\special{pa 4030 1860}%
\special{fp}%
% VECTOR 2 0 3 0
% 2 3960 1180 3960 1160
% 
\special{pn 8}%
\special{pa 3960 1180}%
\special{pa 3960 1160}%
\special{fp}%
\special{sh 1}%
\special{pa 3960 1160}%
\special{pa 3940 1228}%
\special{pa 3960 1214}%
\special{pa 3980 1228}%
\special{pa 3960 1160}%
\special{fp}%
% LINE 2 0 3 0
% 2 3600 600 3600 1920
% 
\special{pn 8}%
\special{pa 3600 600}%
\special{pa 3600 1920}%
\special{fp}%
% STR 2 0 3 0
% 3 4140 1790 4140 1890 5 0
% $\gamma$
\put(41.4000,-18.9000){\makebox(0,0){$\gamma$}}%
% STR 2 0 3 0
% 3 4080 890 4080 990 5 0
% $\Gamma(\gamma)$
\put(40.8000,-9.9000){\makebox(0,0){$\Gamma(\gamma)$}}%
\end{picture}%